# Influence of Dispersity on the Relaxation of Entangled Polymers from Molecular Dynamics Simulations

Taofeek Tejuosho[1], Janani Sampath[1,*]

*jsampath@ufl.edu

[1]Department of Chemical Engineering, University of Florida, Gainesville, Florida 32611

**Abstract**

Nonequilibrium molecular dynamics simulations are used to study the deformation behavior of disperse polymer melts by tracking test chains of length $N = M_w$, the weight average molecular weight, in melts of varying dispersity. At high strain rates, stress decreases with increasing dispersity up to the critical stretch ratio. After flow cessation, relaxation is anisotropic: the parallel component of the radius of gyration decreases monotonically with dispersity, but the perpendicular component exhibits non-monotonic behavior, with fast relaxation at low to intermediate stretching ratios, and slower relaxation at higher stretching rations as dispersity increases. Single-chain structure factor shows a similar trend in the perpendicular direction, with fast relaxation at intermediate length scales in disperse systems, while the parallel component remains unaffected by dispersity. These findings reveal the complex, dispersity-dependent anisotropic relaxation mechanisms in moderately entangled polymer melts far from equilibrium.

Polymer flow is essential for efficient processing via techniques such as film extrusion, melt blowing, fiber spinning. In many of these processes, the polymer is deformed at rates that exceed the characteristic relaxation times of its chains, significantly altering chain conformations far from equilibrium.[1–3] The behavior of polymers under these conditions is complex, with recent studies suggesting that additional physics beyond the reptation model of de Gennes[4] and Doi-Edwards[5], which treats entanglements as temporary constraints that hinder diffusion, is needed to explain behavior of polymer chains in highly nonlinear flows.[6–10] While several studies in recent years have focused on this behavior in monodisperse melts,[11–15] polymer chain dispersity plays a central role in influencing the rheological and mechanical response of polymer melts.[16–18] However, its role in influencing chain relaxation after strain cessation remains underexplored. This represents a significant gap, as industrially synthesized polymers inherently exhibit some degree of dispersity. Graessley and co-workers showed that viscoelastic responses driven by chain dynamics in the melt and their deformation processes are highly sensitive to molecular weight distribution.[19] Recently, Peters et. al. [20,21] performed coarse grained molecular simulations based on a polyethylene model to probe dispersity effects on equilibrium viscoelastic response for narrowly dispersed melt (Đ < 1.16) and found that terminal relaxation time decreases with increasing dispersity, as faster motion of shorter chains result in constraint release for the longer ones in the melt. Given the impact of dispersity on equilibrium behavior of polymer melts, we conduct coarse-grained molecular dynamics simulations to elucidate how dispersity alters relaxation pathways in entangled polymer melts under flow. We explore the effect of chain length dispersity on the conformation and entanglement statistics of test chains (chains of the same $M_w$ but in melts of varying dispersities) in the nonlinear viscoelastic regime, as well as the relaxation after the deformation ceases.

We perform molecular dynamics (MD) simulations of strongly deformed polymer using the GPU-accelerated LAMMPS package.[22] Polymer chains are modeled using the Kremer–Grest bead spring model and follow the Schulz – Zimm molecular weight distribution.[12,15,23] Model details can be found in a prior study[24] and Supporting Information (SI). To simulate uniaxial extensional flow, the simulations box is stretched along the z-axis by a factor of $\lambda(t)$, while the x and y dimensions contract by $\lambda(t)^{-1/2}$ to maintain constant volume, assuming polymer melt incompressibility. We apply varying Hencky strain rates $\dot{\epsilon} = d \ln \lambda/dt$, similar to isochoric elongation protocol used by Hsu and Kremer.[25] Our systems are large enough that even at the highest strain rates considered, we do not observe any deformation induced instabilities.[26] The lateral box dimension at highest stretching ratio of $\lambda = 20$ (Hencky strain ~ 3) is ~18 $\sigma$, greater than the radius of gyration $R_g = 10.4\ \sigma$ for N = 360. Due to this, we do not use the Kraynik-Reinelt boundary conditions,[27] and our results are comparable with prior simulation work on chain relaxation by Hsu and Kremer[25] and Xu et al.[12] The Rouse-Weissenberg number, $Wi_R = \dot{\epsilon}\tau_R$, quantifies the deformation rate to relative chain relaxation, where $\tau_R = \tau_e \langle Z \rangle^2$ is the Rouse time, $\tau_e$ is the entanglement time, and $\langle Z \rangle$ is the average number of entanglements per chain. Previous studies[27] have measured $\tau_e \approx 1.98 \times 10^3$, and our Z1+ analysis for the monodisperse melt ($M_w$ = N = 360) yields $\langle Z \rangle \approx 13$. Simulations are conducted at $Wi_R$ values ranging from 2.5 to 25, which is within $\tau_R^{-1} < \dot{\epsilon} < \tau_e^{-1}$. In this regime, the interplay between the deformation rate and the polymer's internal conformational relaxation becomes important.[28]

The extensional viscosity $\eta_E^+$ as a function of time $t$ for melts of different dispersities at $Wi_R$ = 2.5 and 25 are shown in Figure 1. For $Wi_R = 2.5$ (Figure 1a), although chains still undergo disentanglement during deformation, they have sufficient time to partially relax, redistribute their stress, and restore some entanglements before further stretching, making the response largely independent of melt dispersity. In the case of $Wi_R = 25$, as the deformation progresses, the viscosity rises sharply, indicating the onset of elongation hardening which decreases with increasing Đ (Figure 1b). Also shown in the inset, monodisperse melt exhibits the earliest onset of strain hardening and the steepest initial slope during strain hardening, suggesting more efficient stress transfer along uniformly long chains. Although strain hardening is delayed, the moderately disperse melt (Đ = 1.4) attains a higher stress peak due to the combined effects of delayed relaxation of relatively long chains and rapid recoil of shorter ones. In contrast, the highly disperse melt (Đ = 2.0) shows the lowest stress peak with a gradual rise, indicating a broader range of relaxation times, thereby slowing the overall stress buildup. We attribute this behavior to the increasing concentration of shorter chains, which act as plasticizing agents and facilitate load transfer to the longer chains. We see that disperse melts yield at a higher critical stretching ratio (λ ≈ 17 for Đ = 1.4 and 2.0 vs. λ ≈ 15 for Đ = 1.0). We analyze the total pair distribution function g(r) (shown in Figure S1) at this critical stretching ratios and observe that the increasing peak intensity which reflects stronger interchain packing due to flow alignment, decreases with increasing dispersity. Using the RepTate toolkit,[29] we fit our simulation results to the Rolie-Double-Poly (RDP) model,[30] a tube-based constitutive model developed to predict nonlinear flow conditions in polydisperse melts. Represented by the solid, dashed, and dotted lines for Đ = 1.0, 1.4 and 2.0 respectively in Fig. 1, we fit the model to the evolution of stress data for both rates ($Wi_R = 2.5$ and $25$). We observe that the RDP model can capture the stress responses at low to moderate rates than at high rates, where deviations emerge at high stretching ratios. This discrepancy arises from the model's simplification of chain stretch relaxation as a single mode, which fails to capture the full Rouse spectrum of stretch dynamics. Polymer chains relax through multiple normal modes, each with distinct timescales. At high strain rates, the higher-frequency odd modes are excited, amplifying their role in governing internal chain relaxation—an effect not captured by the RDP model.

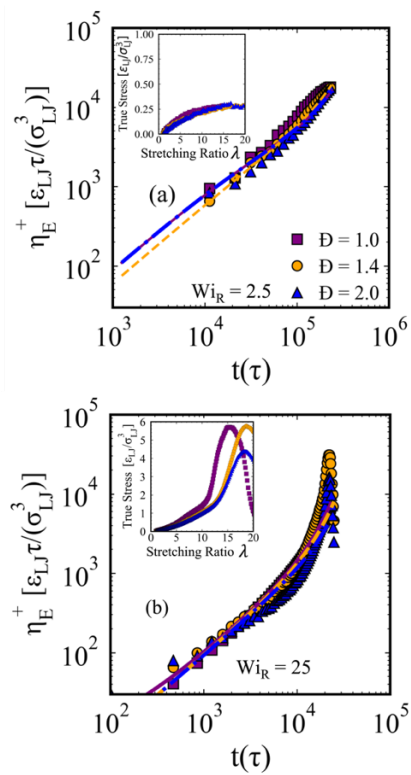

Figure 1. Extensional viscosity $\eta_E^+$ as a function of time for the different melts with $M_w = 360$ at (a) $Wi_R = 2.5$ and (b) $Wi_R = 25$. Insets show engineering stress as a function of stretching ratio $\lambda$. The solid (Đ = 1.0), dashed (Đ = 1.4), and dotted lines (Đ = 2.0) represent the RDP model predictions from RepTate rheology software.[29] Parameters used for fitting can be found in the SI.

Next, we analyze the conformational evolution of test chains ($N = M_W = 360$) within different melts via the radius of gyration as shown in Figure 2a for $Wi_R = 25$. At short and intermediate stretching ratios, the size of the test chain increases, indicating flow-induced stiffening and an increase in persistence length, with the ratio $\langle R_{ee}^2 \rangle/(6\langle R_g^2 \rangle)$ increasing from its equilibrium value of 1 to larger values, as seen in Figure S2 of the SI, consistent with prior findings.[11] Beyond $\lambda = 18$, we observe a clear separation in chain size as a function of dispersity, with the test chains in lower-dispersity melts exhibiting a more compact conformation compared to high dispersity systems, because the presence of a few extremely long, slow-relaxing chains in the disperse melts delays the relaxation of the shorter test chains, leading to a more extended conformation and higher $\langle R_g \rangle$ values at larger deformations. At low to moderate strain rates, chain conformation is indistinguishable across all melt dispersities, as seen in Figure S2. In Figure 2b, we plot the primitive path of the contour length $\langle L_{PP} \rangle$ for test chains as a function of stretching ratio. At high strain rates ($Wi_R = 25$), the evolution of $\langle L_{PP} \rangle$ with stretching ratio increases with decreasing melt dispersity. In monodisperse melts, test chain's $\langle L_{PP} \rangle$ increases and then decreases sharply beyond a critical stretching ratio $\lambda_C = 18$, reflecting a network where all chains contribute almost equally to the topological constraints. In contrast, disperse melts exhibit lower $\langle L_{PP} \rangle$ compared to the monodisperse melt, which is evident even at low stretching ratios. Shorter chains form fewer entanglements with the test chains, while longer ones (N > 360) that are well entangled dominate the stress-bearing network. At the stretching ratios considered, we do not observe a decrease in $\langle L_{PP} \rangle$ for the disperse melts. The loss of topological constraints increases with deformation rate as chains lack sufficient time to interdigitate, consistent with prior studies.[31] More importantly, the extent of disentanglement increases with melt dispersity. In highly disperse melts, the increasing concentration of shorter, more mobile chains surrounding the entangled test chains accelerates entanglement attenuation compared to monodisperse systems, where surrounding chains possess a similar number of entanglements $\langle Z \rangle$, as seen in Figure 2c. Under rapid deformation and at high stretching ratios, there is an increase in the convective constraint release (CCR) mechanism,[2] in which entanglement constraints are released due to the retraction of chain contour length induced by flow. This effect becomes particularly significant for polymer chains in disperse melts.[32]

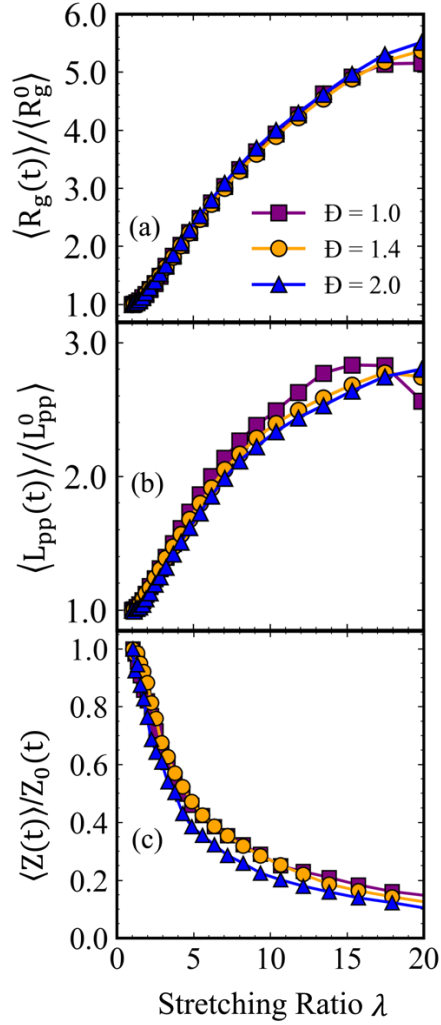

Figure 2. Conformations of test chain of length N = 360 in melts of varying dispersity as function of stretching ratio $\lambda$ for strain rate of $Wi_R = 25$. All melts have the same weight average molecular weight $M_w = 360$. (a) Radius of gyration $\langle R_g \rangle$ (b) Evolution of the contour length of the primitive path $\langle L_{PP} \rangle$ (c) Normalized average number of entanglements $\langle Z \rangle$.

Figure 3a shows a snapshot of the relaxation of randomly chosen test chains in the melt for the different systems, which suggests that test chains in disperse melts tend to recover their equilibrium conformation more readily than chains in the monodisperse melt. We decompose the parallel $\langle R_g^{\parallel} \rangle$ and perpendicular $\langle R_g^{\perp} \rangle$ components of the mean square radius of gyration and capture relaxation mechanisms of these test chains due to the heterogeneity of their surroundings; this is shown in Figures 3b and 3c. We observe that $\langle R_g^{\parallel} \rangle$ decreases and $\langle R_g^{\perp} \rangle$ increases monotonically, approaching their respective equilibrium values. This reflects entropically driven compression ($\langle R_g^{\parallel} \rangle$) and expansion ($\langle R_g^{\perp} \rangle$) during stress relaxation, as chains return to their isotropic random coil conformations. This behavior aligns with previous findings in polymer melts of similar chain lengths.[12,14] For $\langle R_g^{\parallel} \rangle$, we see that at early times, the curves collapse on each other. However, at later times, chain relaxation increases with dispersity, as the presence of shorter, more mobile chains that relax quickly after deformation loosens the entanglement network. As a result, $\langle R_g^{\parallel} \rangle$ at a given time decreases with increasing dispersity. The evolution of $\langle R_g^{\perp} \rangle$ is less straightforward. At early times, dispersity enhances lateral expansion due to a higher proportion of fast-recoiling short chains, with the Đ = 2.0 system exhibiting the fastest convergence toward equilibrium. Meanwhile, the Đ = 1.0 and 1.4 systems display similar $\langle R_g^{\parallel} \rangle$ values due to the lower proportion of short chains in the Đ = 1.4 system, compared to the Đ = 2.0 system. However, relaxation of long entangled chains dominates at later times thereby limiting the expansion of short chains, which leads to a non-monotonic dispersity effect in the direction perpendicular to extension. Particularly, the Đ = 1.4 melt represents a transitional regime where the melt is not dilute enough with short chains to promote fast long-time relaxation, but too heterogeneous to behave like a coherent monodisperse melt, giving rise to a non-monotonic influence of dispersity in the evolution of $\langle R_g^\perp \rangle$. This analysis highlights the role of dispersity in modulating anisotropic conformational recovery.

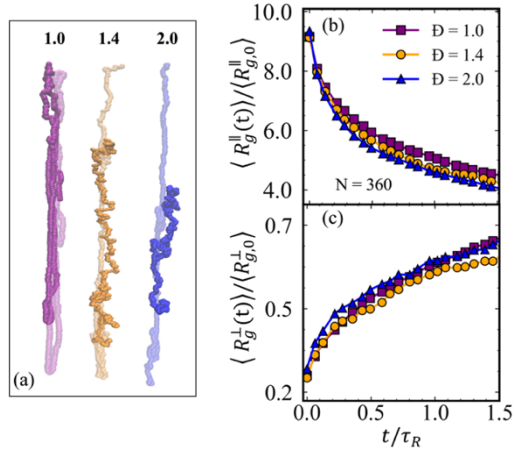

Figure 3. (a) Snapshots of conformational relaxation of randomly chosen test chain in melt of different dispersity at relaxation time $t/\tau_R = 1.5$. Initial state after deformation $t/\tau_R = 0$ is translucent. (b) Evolution of the Radius of gyration in the (b) parallel and (c) perpendicular directions during stress relaxation, normalized by their equilibrium values. Note that these results are after deformation is stopped for a strain rate of $Wi_R = 25$ and stretching ratio $\lambda = 15$.

We compute the normalized single chain structure factor $S(q)$ to capture anisotropy after flow stops over different length scales. Immediately after cessation of deformation, as shown in Figure 4, we see that $S_\perp(q_\perp)$ and $S_\parallel(q_\parallel)$, the components perpendicular and parallel to the stretching direction, respectively, deviate from ideality. As the relaxation time increases, the chains slowly recover ideal behavior, however, even at $t/\tau_R = 1.0$, the chains are significantly far from ideal conformation, with the deviation from ideality decreasing with increasing dispersity. The behavior in the Guinier regime ($q < 2\pi R_g$) is well described by the Debye function, consistent with previous studies.[15] Importantly, both *S(q)* curves collapse on each other at low rates–indicating competing effects between relaxation and deformation rate irrespective of the melt dispersity, at least for up to Đ = 2.0, for all selected relaxation times (Figure S4 in the SI). This means that as test chains are continually stretched, they have enough time to relax back at the same rate for all dispersities. In the case of $Wi_R = 25$, we see that the evolution of $S_\parallel(q_\parallel)$ is independent of dispersity, because test chain relaxation is primarily governed by the test chain's own contour length which does not change with dispersity. In contrast, we see a clear separation between test chains for $S_\perp(q_\perp)$ at high q ($> 2\pi R_g$), as shorter chains in disperse melts facilitate faster lateral relaxation due to reduced confinement. By converting the single chain structure factors into their spherical harmonic representations,[33] we show in Figure S4 of the SI that anisotropy in deformation relaxes monotonically for test chains in melts, regardless of dispersity, as evidenced by the absence of a peak shift up to $2\tau_R$.

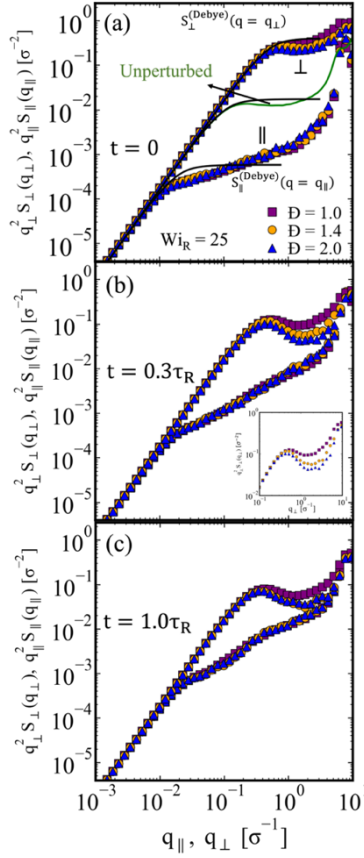

Figure 4. Kratky plot of the two components of the single chain structure factor of test chain of size *N = 360* in melt of different dispersity, parallel and perpendicular to the stretching direction for selected relaxation time. (a) $t/\tau_R = 0$, (b) $t/\tau_R = 0.3$ and (c) $t/\tau_R = 1.0$, respectively for $Wi_R = 25$. Inset in (b) shows distinct separation in the perpendicular direction. Results are after deformation is stopped for stretching ratio *λ = 15*.

Finally, we report the stress relaxation of the melts in its entirety from simulations.[27–29] As shown in Fig. 5a, increasing dispersity leads to faster terminal relaxation time, due to constraint release offered by faster motion of the shorter chains on the longer ones.[21] The inset of the plot highlights deviations from Rouse model, which predicts a $t^{-1/2}$ scaling in stress relaxation, given by $G^{Rouse}(t) = (\sqrt{\pi}/2\sqrt{2})\rho_{chain}k_BT(t/\tau_R)^{-1/2}$ for $(t \lesssim \tau_R)$. Where $\rho_{chain}$ is the number density of chains, related to the bead number density, $\rho_{chain} = \rho_{bead}/N$. Based on N = 360 ($k_\theta$ = 1.5$\varepsilon$), the Rouse prediction of $t^{1/2}G^{Rouse}(t) \approx 0.87$ (in LJ reduced units). We see a Rouse plateau appearing at intermediate times, followed by a spike which is a manifestation of entanglement and eventually reaching an inflection point then a rapid decay. We show that the storage $G'$ and loss $G''$ at low frequencies decrease with increasing dispersity (Fig. S5 of SI). We also analyze the stress relaxation after deformation ceases for $Wi_R = 25$ as seen in Fig. 5b. At short to intermediate times, the stress plateaus before eventually decaying as the polymer chains relax toward their equilibrium configuration. This relaxation rate increases with dispersity, although the separation between the two disperse melts is less pronounced. Even with the presence of chains much longer than the test chains in the monodisperse melt, the effects of shorter chains in the disperse melts offer some constraint release leading to faster stress relaxation, similar to the equilibrium results.

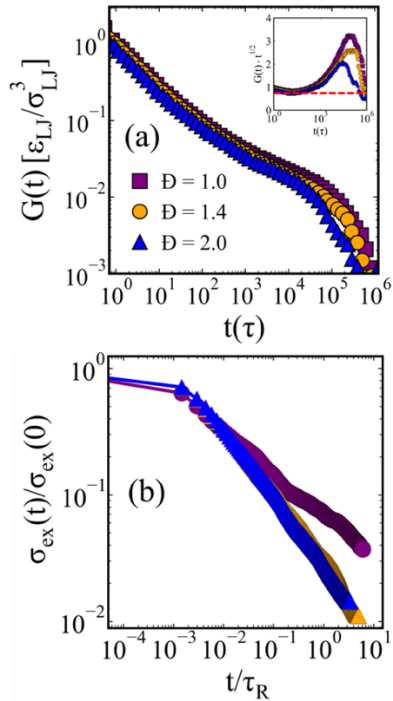

Figure 5. (a) Stress relaxation modulus G(t) of polymer melts of melt of $M_w = 360$ for different dispersity Đ. Inset of the plot showing G(t) scaled by $t^{1/2}$, Horizontal red dashed line shows Rouse model prediction. (b) Stress relaxation of the deformed polymer melt at $\lambda \approx 15$, following cessation of deformation. $\sigma_{ex} = \sigma_{zz} - (\sigma_{xx} + \sigma_{yy})/2$ is the tensile stress, defined as the difference in the diagonal components of the stress tensor $\sigma_{\alpha\beta}$.

In summary, molecular simulations reveal that stresses in deformed polymer melts is largely independent of dispersity at low strain rates but has a non-monotonic dependence on dispersity at high rates, where pronounced strain hardening emerges. This behavior is driven by the enhanced mobility of shorter chains in disperse melts, leading to greater disentanglement. Comparison with the Rolie-Double-Poly (RDP) constitutive model shows good agreement at low to moderate strain rates, but deviations arise under strong flow at longer timescales, corresponding to greater chain stretching. Notably, we observe a complex relaxation landscape in disperse melts, arising from heterogeneous dynamics imposed by chains of varying lengths. This work underscores the critical role of chain length dispersity—a hallmark of industrial polymers—in shaping out-of-equilibrium behavior under strong flow. While we consider moderately entangled melts here, we anticipate these trends will be more pronounced at high molecular weights. Our results can help improve constitutive models for polydisperse polymers at high strain rates and guide the design of materials that are both processable and mechanically robust without altering chemistry.

**Data Availability Statement**

Analysis codes, Lammps input script, python script to create initial configuration of the systems can be found here – [Github Page - Analysis Codes and Lammps Scripts](#)

**Supporting Information (SI)**

Supplementary data is available free of charge

**AUTHOR INFORMATION**

**Corresponding Author**

**Janani Sampath** – Department of Chemical Engineering, University of Florida, Gainesville, Florida 32611, United States

https://orcid.org/0000-0001-6005-9096

Email: jsampath@ufl.edu

**Authors**

**Taofeek Tejuosho** – Department of Chemical Engineering, University of Florida, Gainesville, Florida 32611, United States

https://orcid.org/0009-0009-2186-7651

**Conflicts of Interest**

The authors declare no conflicting interests.

**Acknowledgements**

This work was supported by the donors of ACS Petroleum Research Fund under Doctoral New Investigator Grant 67036-DNI6. The authors acknowledge University of Florida Research

Computing for providing computing resources and support that have contributed to the research results reported in this publication; https://rc.ufl.edu/. We thank Dr. Daniel Read for insights regarding the Rolie-Double-Poly model parameters and Dr. Sachin Shanbhag for many helpful comments about our results.**References**

(1)   M. Doi and S. Edwards. *The Theory of Polymer Dynamics (Oxford University Press)*; 1986.

(2)   Graham, R. S.; Likhtman, A. E.; McLeish, T. C. B.; Milner, S. T. Microscopic Theory of Linear, Entangled Polymer Chains under Rapid Deformation Including Chain Stretch and Convective Constraint Release. *J Rheol (N Y N Y)* **2003**, *47* (5), 1171–1200. DOI: 10.1122/1.1595099.

(3)   Arrigo, R.; Malucelli, G.; Mantia, F. P. L. Effect of the Elongational Flow on the Morphology and Properties of Polymer Systems: A Brief Review. *Polymers (Basel)* **2021**, *13* (20). DOI: 10.3390/polym13203529.

(4)   de Gennes, P. G. Reptation of a Polymer Chain in the Presence of Fixed Obstacles. *J. Chem. Phys.* **1971**, *55* (2), 572–579. DOI: 10.1063/1.1675789.

(5)   de Gennes, P. G. *Scaling Concepts in Polymer Physics*. *Phys. Today* **1979**. DOI: 10.1063/1.2914118.15

(6)     Bach, A.; Almdal, K.; Rasmussen, H. K.; Hassager, O. Elongational Viscosity of Narrow Molar Mass Distribution Polystyrene. *Macromolecules* **2003**, *36* (14), 5174–5179. DOI: 10.1021/ma034279q.

(7)     Wingstrand, S. L.; Alvarez, N. J.; Huang, Q.; Hassager, O. Linear and Nonlinear Universality in the Rheology of Polymer Melts and Solutions. *Phys. Rev. Lett.* **2015**, *115* (7), 078302. DOI: 10.1103/PhysRevLett.115.078302.

(8)     Huang, Q.; Alvarez, N. J.; Matsumiya, Y.; Rasmussen, H. K.; Watanabe, H.; Hassager, O. Extensional Rheology of Entangled Polystyrene Solutions Suggests Importance of Nematic Interactions. *ACS Macro Lett.* **2013**, *2* (8), 741–744. DOI: 10.1021/mz400319v.

(9)     Taghipour, H.; Hawke, L. G. D. Entangled Linear Polymers in Fast Shear and Extensional Flows: Evaluating the Performance of the Rolie-Poly Model. *Rheol. Acta* **2021**, *60* (10), 617–641. DOI: 10.1007/s00397-021-01295-z.

(10)    Huang, Q. When Polymer Chains Are Highly Aligned: A Perspective on Extensional Rheology. *Macromolecules* **2022**, *55* (3), 715–727. DOI: 10.1021/acs.macromol.1c02262.

(11)    Baig, C.; Mavrantzas, V. G.; Kröger, M. Flow Effects on Melt Structure and Entanglement Network of Linear Polymers: Results from a Nonequilibrium Molecular Dynamics Simulation Study of a Polyethylene Melt in Steady Shear. *Macromolecules* **2010**, *43* (16), 6886–6902. DOI: 10.1021/ma100826u.

(12)    Xu, W.-S.; Carrillo, J.-M. Y.; Lam, C. N.; Sumpter, B. G.; Wang, Y. Molecular Dynamics Investigation of the Relaxation Mechanism of Entangled Polymers after a Large Step Deformation. *ACS Macro Lett.* **2018**, *7* (2), 190–195. DOI: 10.1021/acsmacrolett.7b00900.

(13)    Blanchard, A.; Graham, R. S.; Heinrich, M.; Pyckhout-Hintzen, W.; Rciher, D.; Likhtman, A. E.; McLeish, T. C. B.; Read, D. J.; Straube, E.; Kohlbrecher, J. Small Angle Neutron Scattering Observation of Chain Retraction after a Large Step Deformation. *Phys. Rev. Lett.* **2005**, *95* (16), 166001. DOI: 10.1103/PhysRevLett.95.166001.

(14)    Wang, Z.; Lam, C. N.; Chen, W.-R.; Wang, W.; Liu, J.; Liu, Y.; Porcar, L.; Stanley, C. B.; Zhao, Z.; Hong, K.; Wang, Y. Fingerprinting Molecular Relaxation in Deformed Polymers. *Phys. Rev. X* **2017**, *7* (3), 031003. DOI: 10.1103/PhysRevX.7.031003.

(15)    Hsu, H.-P.; Kremer, K. Chain Retraction in Highly Entangled Stretched Polymer Melts. *Phys. Rev. Lett.* **2018**, *121* (16), 167801. DOI: 10.1103/PhysRevLett.121.167801.

(16)    Sifri, R. J.; Padilla-Vélez, O.; Coates, G. W.; Fors, B. P. Controlling the Shape of Molecular Weight Distributions in Coordination Polymerization and Its Impact on Physical Properties. *J. Am. Chem. Soc.* **2020**, *142* (3), 1443–1448. DOI: 10.1021/jacs.9b11462.

(17)   Whitfield, R.; Truong, N. P.; Anastasaki, A. Precise Control of Both Dispersity and Molecular Weight Distribution Shape by Polymer Blending. *Angew. Chem. Int. Ed* **2021**, *60* (35), 19383–19388. DOI: 10.1002/anie.202106729.

(18)   Gentekos, D. T.; Dupuis, L. N.; Fors, B. P. Beyond Dispersity: Deterministic Control of Polymer Molecular Weight Distribution. *J. Am. Chem. Soc.* **2016**, *138* (6), 1848–1851. DOI: 10.1021/jacs.5b13565.

(19)   Wasserman, S. H.; Graessley, W. W. Effects of Polydispersity on Linear Viscoelasticity in Entangled Polymer Melts. *J Rheol (N Y N Y)* **1992**, *36* (4), 543–572. DOI: 10.1122/1.550363.

(20)   Peters, B. L.; Salerno, K. M.; Ge, T.; Perahia, D.; Grest, G. S. Effect of Chain Length Dispersity on the Mobility of Entangled Polymers. *Phys. Rev. Lett.* **2018**, *121* (5), 057802. DOI: 10.1103/PhysRevLett.121.057802.

(21)   Peters, B. L.; Salerno, K. M.; Ge, T.; Perahia, D.; Grest, G. S. Viscoelastic Response of Dispersed Entangled Polymer Melts. *Macromolecules* **2020**, *53* (19), 8400–8405. DOI: 10.1021/acs.macromol.0c01403.

(22)   Thompson, A. P.; Aktulga, H. M.; Berger, R.; Bolintineanu, D. S.; Brown, W. M.; Crozier, P. S.; in 't Veld, P. J.; Kohlmeyer, A.; Moore, S. G.; Nguyen, T. D.; Shan, R.; Stevens, M. J.; Tranchida, J.; Trott, C.; Plimpton, S. J. LAMMPS - a Flexible Simulation Tool for Particle-Based Materials Modeling at the Atomic, Meso, and Continuum Scales. *Comput. Phys. Commun.* **2022**, *271*, 108171. DOI: 10.1016/j.cpc.2021.108171.

(23)   Kremer, K.; Grest, G. S. Dynamics of Entangled Linear Polymer Melts:  A Molecular-dynamics Simulation. *J. Chem. Phys.* **1990**, *92* (8), 5057–5086. DOI: 10.1063/1.458541.

(24)   Tejuosho, T.; Kollipara, S.; Patankar, S.; Sampath, J. Dynamics of Polymer Chains in Disperse Melts: Insights from Coarse-Grained Molecular Dynamics Simulations. *J. Phys. Chem. B* **2024**. DOI: 10.1021/acs.jpcb.4c05610.

(25)   Hsu, H.-P.; Kremer, K. Primitive Path Analysis and Stress Distribution in Highly Strained Macromolecules. *ACS Macro Lett.* **2018**, *7* (1), 107–111. DOI: 10.1021/acsmacrolett.7b00808.

(26)   Zheng, Y.; Tsige, M.; Wang, S.-Q. Molecular Dynamics Simulation of Entangled Melts at High Rates: Identifying Entanglement Lockup Mechanism Leading to True Strain Hardening. *Macromol. Rapid Commun.* **2023**, *44* (1), e2200159. DOI: 10.1002/marc.202200159.

(27)   O'Connor, T. C.; Alvarez, N. J.; Robbins, M. O. Relating Chain Conformations to Extensional Stress in Entangled Polymer Melts. *Phys. Rev. Lett.* **2018**, *121* (4), 047801. DOI: 10.1103/PhysRevLett.121.047801.

(28) Likhtman, A. E.; McLeish, T. C. B. Quantitative Theory for Linear Dynamics of Linear Entangled Polymers. *Macromolecules* **2002**, *35* (16), 6332–6343. DOI: 10.1021/ma0200219.

(29) Boudara, V. A. H.; Read, D. J.; Ramírez, J. Reptate Rheology Software: Toolkit for the Analysis of Theories and Experiments. *J Rheol (N Y N Y)* **2020**, *64* (3), 709–722. DOI: 10.1122/8.0000002.

(30) Boudara, V. A. H.; Peterson, J. D.; Leal, L. G.; Read, D. J. Nonlinear Rheology of Polydisperse Blends of Entangled Linear Polymers: Rolie-Double-Poly Models. *J Rheol (N Y N Y)* **2019**, *63* (1), 71–91. DOI: 10.1122/1.5052320.

(31) Song, J.; Li, J.; Li, Z. Molecular Dynamics Simulations of Uniaxial Deformation of Bimodal Polyethylene Melts. *Polymer* **2021**, *213*, 123210. DOI: 10.1016/j.polymer.2020.123210.

(32) Mead, D. W.; Larson, R. G.; Doi, M. A Molecular Theory for Fast Flows of Entangled Polymers. *Macromolecules* **1998**, *31* (22), 7895–7914. DOI: 10.1021/ma980127x.

(33) Viovy, J. L. Chain Retraction, Length Fluctuations, and Small-angle Neutron Scattering in Strained Polymer Melts. *J. Polym. Sci. B Polym. Phys.* **1986**, *24* (7), 1611–1618. DOI: 10.1002/polb.1986.090240717. Sciwheel inserting bibliography...Sciwheel inserting bibliography...